\documentclass[journal]{IEEEtran}
\usepackage{fancyhdr}

\usepackage[acronyms,nonumberlist,nopostdot,nomain,nogroupskip]{glossaries}
\newacronym{6g}{6G}{sixth generation}
\newacronym{5g}{5G}{fifth generation}
\newacronym{snr}{SNR}{signal-to-noise ratio}
\newacronym{bw}{BW}{bandwidth}
\newacronym[plural=\gls{cnn}s,firstplural=convolutional neural networks (CNNs)]{cnn}{CNN}{convolutional neural network}

\newacronym{iq}{I/Q}{in phase/quadrature}
\newacronym{ml}{ML}{machine learning}
\newacronym{phy}{PHY}{physical}
\newacronym[plural=\gls{dnn}s,firstplural=deep neural networks (DNNs)]{dnn}{DNN}{deep neural network}
\newacronym{mmwave}{mmWave}{millimeter wave}
\newacronym{dsp}{DSP}{digital signal processing}
\newacronym{dsa}{DSA}{dynamic spectrum access}
\newacronym{ism}{ISM}{industrial, scientific and medical}
\newacronym{csi}{CSI}{channel state information}
\newacronym{fcc}{FCC}{Federal Communication Commission}
\newacronym{rfp}{RFP}{radio fingerprinting}
\newacronym[plural=\gls{sdr}s,firstplural=long training fields (SDRs)]{sdr}{SDR}{software-defined radio}

\newacronym{iot}{IoT}{Internet of things}
\newacronym{mimo}{MIMO}{multi-input, multi-output}
\newacronym{mum}{MU-MIMO}{multi-user MIMO}
\newacronym{sum}{SU-MIMO}{single-user multi-input, multi-output}
\newacronym{iui}{IUI}{inter-user interference}
\newacronym{isi}{ISI}{inter-stream interference}

\newacronym[plural=\gls{wlan}s,firstplural=wireless local area networks (WLANs)]{wlan}{WLAN}{wireless local area network}

\newacronym{ap}{AP}{access point}
\newacronym{sta}{STA}{station}
\newacronym{dl}{DL}{downlink}

\newacronym{mcs}{MCS}{modulation and coding scheme}
\newacronym{cfr}{CFR}{channel frequency response}
\newacronym{cir}{CIR}{channel impulse response}
\newacronym{ndp}{NDP}{null data packet}
\newacronym[plural=\gls{ltf}s,firstplural=long training fields (LTFs)]{ltf}{LTF}{long training field}
\newacronym{vht}{VHT}{very high throughput}
\newacronym{ht}{HT}{high throughput}

\newacronym{ofdm}{OFDM}{orthogonal frequency-division multiplexing}
\newacronym{ofdma}{OFDMA}{orthogonal frequency-division multiple access}
\newacronym{cfo}{CFO}{carrier frequency offset}
\newacronym{sfo}{SFO}{sampling frequency offset}
\newacronym{pdd}{PDD}{packet detection delay}
\newacronym{ppo}{PPO}{phase-locked loop offset}
\newacronym{pll}{PLL}{phase-locked loop}
\newacronym{pa}{PA}{phase ambiguity}
\newacronym{sbc}{SBC}{single board computer}

\newacronym[plural=\gls{cm}s,firstplural=confusion matrices (CMs)]{cm}{CM}{confusion matrix}

\newacronym{id}{ID}{identifier}
\newacronym{aoa}{AoA}{angle of arrival}
\newacronym{ul}{UL}{uplink}

\newacronym{svd}{SVD}{singular value decomposition}

\newacronym[plural=\gls{pdf}s,firstplural=probability density functions (PDFs)]{pdf}{PDF}{probability density function}

\newacronym{har}{HAR}{human activity recognition}  
\newacronym{cots}{COTS}{commercial-off-the-shelf} 
\newacronym[plural=\gls{imu}s,firstplural=inertial measurement units (IMUs)]{imu}{IMU}{inertial measurement unit}
\newacronym[plural=\gls{usrp}s,firstplural=universal software radio peripherals (USRPs)]{usrp}{USRP}{universal software radio peripheral}
\newacronym[plural=\gls{svm}s,firstplural=support vector machines (SVMs)]{svm}{SVM}{support vector machine}
\newacronym{ifft}{IFFT}{inverse fast Fourier transform}
\newacronym{fmcw}{FMCW}{frequency-modulated continuous-wave}
\newacronym[plural=\gls{enb}s,firstplural= evolved nodes base (eNBs)]{enb}{eNB}{evolved node base}

\newacronym{mac}{MAC}{medium access control}

\newacronym{he}{HE}{high-efficiency}
\newacronym{eht}{EHT}{extremely high throughput} 
\newacronym{ng60}{NG60}{next generation 60~GHz} 
\newacronym{3gpp}{3GPP}{Third Generation Partnership Project} 
\newacronym{nr}{NR}{new radio}
\newacronym{nru}{NR-U}{new radio-based access to unlicensed spectrum}
\newacronym{ieee}{IEEE}{Institute of Electrical and Electronics Engineers}
\newacronym{gsm}{GSM}{global system for mobile communications}
\newacronym{etsi}{ETSI}{European Telecommunications Standards Institute}
\newacronym{umts}{UMTS}{universal mobile telecommunications system}
\newacronym{lte}{LTE}{long term evolution of UMTS}
\newacronym{itu}{ITU}{International Telecommunications Union}
\newacronym{imt}{IMT}{International Mobile Telecommunications}
\newacronym{scfdma}{SC-FDMA}{single carrier - frequency division multiple access}
\newacronym{qos}{QoS}{quality of service}
\newacronym{qoe}{QoE}{quality of experience}

\newacronym{mmtc}{mMTC}{massive machine type communications}
\newacronym{urllc}{URLLC}{ultra-reliable and low latency communications}
\newacronym{embb}{eMBB}{enhanced mobile broadband}

\newacronym{mec}{MEC}{multi-access edge computing}
\newacronym[plural=\gls{meh}s,firstplural=MEC hosts (MEHs)]{meh}{MEH}{MEC host}

\newacronym{gnss}{GNSS}{global navigation satellite system}

\newacronym[plural=\gls{mc}s,firstplural=Markov chains (MCs)]{mc}{MC}{Markov chain} 
\newacronym[plural=\gls{nn}s,firstplural=neural networks (NNs)]{nn}{NN}{neural network}
\newacronym[plural=\gls{rnn}s,firstplural=recurrent neural networks (RNNs)]{rnn}{RNN}{recurrent neural network}
\newacronym[plural=\gls{gru}s,firstplural=gated recurrent units (GRUs)]{gru}{GRU}{gated recurrent unit}

\newacronym{toa}{ToA}{time of arrival}
\newacronym{tdoa}{TDoA}{time difference of arrival}
\newacronym{rss}{RSS}{received signal strength}
\newacronym{rssi}{RSSI}{received signal strength indicator}
\newacronym{sumo}{SUMO}{simulation of urban mobility}
\newacronym{pl}{PL}{path loss} 
\newacronym{nlos}{NLoS}{non line of sight} 
\newacronym{hdbscan}{HDBSCAN}{hierarchical density-based spatial clustering of applications with noise}

\newacronym{ue}{UE}{user equipment}
\newacronym{iov}{IoV}{Internet of vehicles}
\newacronym[plural=\gls{vm}s,firstplural=virtual machines (VMs)]{vm}{VM}{virtual machine}

\newacronym[plural=\gls{pv}s,firstplural=photovoltaic panels (PVs)]{pv}{PV}{photovoltaic panel}
\newacronym{mpc}{MPC}{model predictive control}
\newacronym{cpu}{CPU}{central processing unit}

\newacronym{qp}{QP}{quadratic programming}  
\newacronym{eh}{EH}{energy harvesting}
\newacronym{ups}{UPS}{uninterrupted power supply}
\newacronym{mcc}{MCC}{mobile cloud computing}
\newacronym{mip}{MIP}{mixed integer programming}
\newacronym{m2m}{M2M}{\mbox{machine-to-machine}}

\newacronym{wg}{WG}{Working Group}
\newacronym{tg}{TG}{Task Group}
\newacronym{par}{PAR}{Project Authorization Request}
\newacronym{frd}{FRD}{Functional Requirement Document}
\newacronym{bf}{TGbf}{IEEE 802.11bf}
\newacronym{sc}{SC}{Standards Committee}
\newacronym{ec}{EC}{Executive Committee}
\newacronym{sfd}{SFD}{Specification Framework Document}
\newacronym{ppdu}{PPDU}{\gls{phy} Protocol Data Unit}
\newacronym{sp}{S\&P}{security and privacy}
\newacronym{mmw}{mmWave}{millimeter wave}

\newacronym[plural=\gls{nap}s,firstplural=non-AP stations (non-AP STAs)]{nap}{non-AP STA}{non-AP station}

\newacronym[plural=\gls{ru}s,firstplural=resource units (RUs)]{ru}{RU}{resource unit}
\usepackage{amsmath,amsfonts}
\usepackage{algorithmic}
\usepackage{algorithm}
\usepackage{array}
\usepackage[caption=false,font=normalsize,labelfont=sf,textfont=sf]{subfig}
\usepackage{textcomp}
\usepackage{stfloats}
\usepackage{url}
\usepackage{verbatim}
\usepackage{graphicx}
\usepackage{cite}
\usepackage{xcolor}

\begin{document}

\title{Toward Integrated Sensing and Communications\\in IEEE 802.11bf Wi-Fi Networks}

\author{\IEEEauthorblockN{Francesca Meneghello$^*$, Cheng Chen$^{\mathsection}$, Carlos Cordeiro$^{\mathsection}$, and Francesco Restuccia$^{\dag}$}\\
 \IEEEauthorblockA{$^*$ Department of Information Engineering, University of Padova, Italy}\\
 \IEEEauthorblockA{$^{\mathsection}$Intel Corporation, United States}\\
 \IEEEauthorblockA{$^{\dag}$ Institute for the Wireless Internet of Things, Northeastern University, United States}\vspace{-0.6cm}
 }

\maketitle

\begin{abstract}
As Wi-Fi becomes ubiquitous in public and private spaces, it becomes natural to leverage its intrinsic ability to sense the surrounding environment to implement groundbreaking wireless sensing applications such as human presence detection, activity recognition, and object tracking. For this reason, the IEEE 802.11bf Task Group is defining the appropriate modifications to existing Wi-Fi standards to enhance sensing capabilities through 802.11-compliant devices. However, the new standard is expected to leave the specific sensing algorithms open to implementation. To fill this gap, this article explores the practical implications of integrating sensing into \mbox{Wi-Fi} networks. We provide an overview of the physical and medium access control layers sensing enablers, together with the application layer perspective. We analyze the impact of communication parameters on sensing performance and detail the main research challenges. To make our evaluation replicable, we pledge to release all of our dataset and code to the community.\footnote{The dataset and code are available at \url{https://github.com/francescamen/SHARPax}}
\end{abstract}
\thispagestyle{fancy}


\section{Introduction}
\thispagestyle{fancy}
\IEEEPARstart{I}n 1997, the \gls{ieee} released the first 802.11 standard. The document specified the \gls{phy} and \gls{mac} layers for wireless local area networks operating on the \textit{unlicensed} portion of the radio spectrum. The name \textit{Wi-Fi} was introduced in 1999 by the Wi-Fi alliance, which ensures interoperability among IEEE 802.11 devices.
Today, Wi-Fi networks are used to connect hundreds of millions of people worldwide. Thus, the research community has suggested leveraging their ubiquitousness for \textit{wireless sensing} applications. This entails obtaining information about objects or people in the environment as they act as radio signals reflectors, diffractors, and/or scatterers, by tracking changes in quantities that describe the way radio signals propagate in the environment. Such quantities are continuously estimated by Wi-Fi devices for communication purposes to properly transmit and decode data. The main idea behind \textit{\mbox{Wi-Fi} sensing} is to use them as \textit{sensing primitives}.
This way, \mbox{Wi-Fi} devices can act as sensors, opening up a plethora of new applications such as human activity and pose recognition, person identification, and the Metaverse, among others~\cite{ma2019wifi,Liu2022Integrated}.

To make Wi-Fi sensing available to the general public, researchers are currently following two parallel and equally important directions. On the one hand, sensing primitives are being made available outside of the communication procedure through the definition of the new IEEE 802.11bf standard, which is expected to be finalized by 2024~\cite{Chen2022WiFi}. On the other hand, researchers are developing accurate and robust sensing algorithms that leverage \mbox{Wi-Fi} sensing primitives. This article aims to bridge these two research lines, providing a vision of the \mbox{Wi-Fi} features -- at the lower layers of the protocol stack -- that are key enablers for sensing (Section~\ref{sec:networks}), and how they can be leveraged to design sensing applications (Section~\ref{sec:algorithms}). Practical suggestions attained from experimental evaluations with commercial IEEE 802.11ax devices and an overview of the research challenges are presented in Sections~\ref{sec:evaluation}-\ref{sec:challenges}.
To the best of our knowledge, no work in the literature provides a holistic view of sensing in Wi-Fi networks. Moreover, this is the first time data from commercial 802.11ax-compliant devices is considered for sensing purposes. In turn, the analysis in Section~\ref{sec:evaluation} is the first to consider the new \gls{ofdma} modulation scheme that has been introduced with 802.11ax and will be adopted also in next-generation 802.11be networks \cite{Chen2022Overview}. \vspace{-0.3cm}

\subsection{The Integrated Sensing and Communications Paradigm}
Sensing operations are set to coexist with data transmissions in upcoming Wi-Fi networks~\cite{Liu2022Integrated}. This concept is usually referred to as \textit{integrated sensing and communications} (in short, ISAC). In addition to Wi-Fi, ISAC is being explored in other radio technologies. The main approaches are \textit{communication-centric} and \textit{sensing-centric}. While the focus of the former is on reusing communication signals for sensing, the latter aims to transmit information through radar-like waveforms~\cite{Zhang2021Overview}. In turn, when accurate sensing measurements are needed for, e.g., safety-critical applications, sensing-centric approaches should be adopted. Instead, when the purpose is to provide sensing functionalities without the burden of installing additional hardware, communication-centric strategies are preferred. The choice also depends on whether the sensing happens indoor or outdoor. In this article, we focus on an indoor communication-centric scenario, while we refer the reader to~\cite{Liu2022Integrated} for an overview of ISAC in next-generation cellular networks.\vspace{-0.2cm}

\section{Physical and Medium Access Control Layers: Wi-Fi Sensing Enablers}\label{sec:networks}
As sensing primitive, most research activities focus on the \gls{csi}, which captures information about the signal \textit{multi-path propagation}. 
Since multi-path is caused by reflections, diffraction, and scattering associated with objects in the environment, it contains rich information for sensing purposes.
The \gls{csi} usually refers to the \gls{cfr}, which is the frequency representation of \gls{cir} (i.e., the time series containing the delay and amplitude of the different paths). The \gls{csi} is continuously estimated for equalization purposes, by leveraging training fields in the data packets~\cite{ma2019wifi}. However, current Wi-Fi standards are designed for communications and do not provide the proper support for the integration of sensing functionalities. Moreover, sensing primitives are not released by commercial devices. Thus, researchers currently leverage ad-hoc procedures to extract CSI. This ultimately hinders the development and commercialization of sensing systems. For this reason, a new IEEE \gls{tg} -- called 802.11bf -- is defining modifications to the 802.11 standards at both the \gls{mac} and \gls{phy} layers to support sensing. The amendment will define a unified procedure to directly obtain the sensing primitives in both the sub-7~GHz and the \gls{mmw} bands. The procedure will involve different devices taking the roles of \textit{initiator}, \textit{responder}, \textit{transmitter}, and \textit{receiver}. The sensing can be \textit{monostatic}, \textit{bistatic}, or \textit{multistatic} based on whether the sensing transmitter and receiver are distinct devices or are the same entity. We refer the reader to \cite{Chen2022WiFi} for a more in-depth overview of IEEE 802.11bf.  \vspace{-0.2cm}

\subsection{Frequency and Spatial Diversity for Sensing}\label{subsec:freq_space_div}

The structure of the \gls{csi} depends on the specific waveform employed. However, an important element to consider is to maximize the diversity that the communication system supports. Indeed, concurrently obtaining data about the propagation of radio waves characterized by different carrier frequencies ({\it frequency diversity}), or captured at different points in space ({\it space diversity}) is crucial to provide good adaptation of the sensing algorithms to changing conditions. 

As a source of frequency diversity, sensing algorithms can leverage the \gls{ofdm} and \gls{ofdma} modulation schemes adopted by \mbox{Wi-Fi} devices (IEEE 802.11n/ac/ax/be/ay). 
Such schemes transmit data over frequency-orthogonal radio spectrum sub-channels. Thus, the per sub-channel \gls{cfr} can be used for sensing purposes. By considering different sub-channels, sensing algorithms can obtain more fine grained ranging information. Another source of frequency diversity resides in simultaneously obtaining data from multiple transmissions in the 2.4 - 7.125~GHz range, and/or in the 57.24 - 70.20~GHz range. Very recently, the \gls{fcc} and the European Commission have opened, respectively, the 5.925-7.125~GHz and 5.945-6.425~MHz spectrum for unlicensed use~\cite{WiFi6Ecountries}.
Spectrum bands above 57 GHz -- \gls{mmw}, used by IEEE 802.11ad/ay devices -- are more challenging from a communication standpoint, yet are appealing for sensing purposes, as they offer wider bandwidths and, in turn, more sensing granularity. 

Spatial diversity can be obtained by leveraging \gls{mimo} and/or performing cooperative sensing. As for the former, since Wi-Fi devices need to obtain the channel information between each pair of transmitter and receiver antennas, data associated with different physical channels can be obtained for sensing purposes.
Cooperative sensing is another way to incorporate spatial diversity into sensing procedures by combining the channel information from multiple \mbox{Wi-Fi} devices. However, this requires strict coordination among the sensing devices to obtain synchronized data starting from the device-specific transmission and collection schedules~\cite{cominelli2023exposing}. 

We remark that IEEE 802.11bf is not expected to specify novel transmission schemes for radio signals. Other IEEE \gls{tg}s such as IEEE 802.11be~\cite{Chen2022Overview} are working on such aspects.

\section{Application Layer: Wi-Fi Sensing Algorithms}\label{sec:algorithms}

While providing the proper support for sensing at the physical and medium access control layers, IEEE 802.11bf is not expected to define specific sensing algorithms. Conversely, the sensing primitives -- collected by leveraging the diversity at the \gls{phy} and \gls{mac} layers -- allow designing sensing applications~\cite{ma2019wifi}. Current approaches can be categorized into {\it model-based}, {\it learning-based}, and {\it hybrid}, as discussed next~\cite{Wu2017Device}. %\vspace{-0.4cm}

\subsection{Model-based Approaches}

This strategy leverages radio propagation models to capture channel variations due to the presence/movement of objects and individuals. Model-based algorithms can be used for example to detect the presence of an object or a person by monitoring the range, Doppler, and angles spectra~\cite{Xiao2022WiSion}. 

The frequency diversity provided by \gls{ofdm} and \gls{ofdma} allows computing the distance between the device and the obstacle in the environment. This is obtained by computing the signal spectrum over the different \gls{ofdm}/\gls{ofdma} sub-channels for each \gls{cfr} estimate. Depending on the length of the propagation path, each copy of the transmitted signal is affected by a time delay that reflects on a frequency shift on each \gls{ofdm}/\gls{ofdma} sub-channel. Therefore, peaks on the spectrum reveal the presence of obstacles and their range. Notice that the range granularity is inversely proportional to the bandwidth. For example, with 160~MHz bandwidth (802.11ax), the range granularity is about 2 meters. In this respect, the newly available 6 GHz and \gls{mmw} bands will be more beneficial for ranging purposes as they provide higher bandwidths~\cite{Milani2018WiFi}. 

The moving velocity of the sensing target can instead be estimated considering the Doppler shift induced by the movements. The estimate is obtained by computing the spectrum over subsequent transmissions with fixed inter-packet time, considering one single \gls{ofdm}/\gls{ofdma} sub-channel. The estimate captures how the frequency shifts associated with the path length vary in time, and thus, the target moving velocity. The results on the available sub-channels can be combined to increase the accuracy of the estimate~\cite{Xie2019MDTrack}.

Spatial diversity allows identifying the angular position of the target by analyzing the phase shift among the signal copies received at the different antennas. The higher the number of antennas, the higher the angular shift granularity~\cite{Xie2019MDTrack}. %\vspace{-0.3cm}

\subsection{Learning-based and Hybrid Approaches}

In general, model-based approaches do not perform well when the sensing task requires recognizing a high number of different situations, e.g., human activities/gestures, with a significant number of activities, and they do not generalize well to multiple subjects and environments. Learning-based approaches, instead, allow capturing more fine-grained features without requiring manual feature extraction from the CSI~\cite{Wu2017Device}. Learning-based techniques span from traditional machine learning algorithms, such as clustering, to advanced deep learning strategies, such as residual networks and attention mechanisms. Hybrid approaches are currently being investigated to leverage the advantages of learning-based and model-based approaches~\cite{Wang2022integrated}. We point out that training learning-based and hybrid techniques require large datasets featuring significant diversity in terms of days of measurements, environments, Wi-Fi hardware, and subjects (in the case of human sensing). This is key to designing algorithms that can generalize well over different domains, thus enabling their implementation on commercial devices for plug-and-play sensing solutions~\cite{Meneghello2022SHARP}. %\vspace{-0.3cm}

\section{Do Communications Parameters Matter? Evaluation with Commercial 802.11 Devices}\label{sec:evaluation}

\begin{figure}[!t]
    \centering
    \includegraphics[width=1\linewidth]{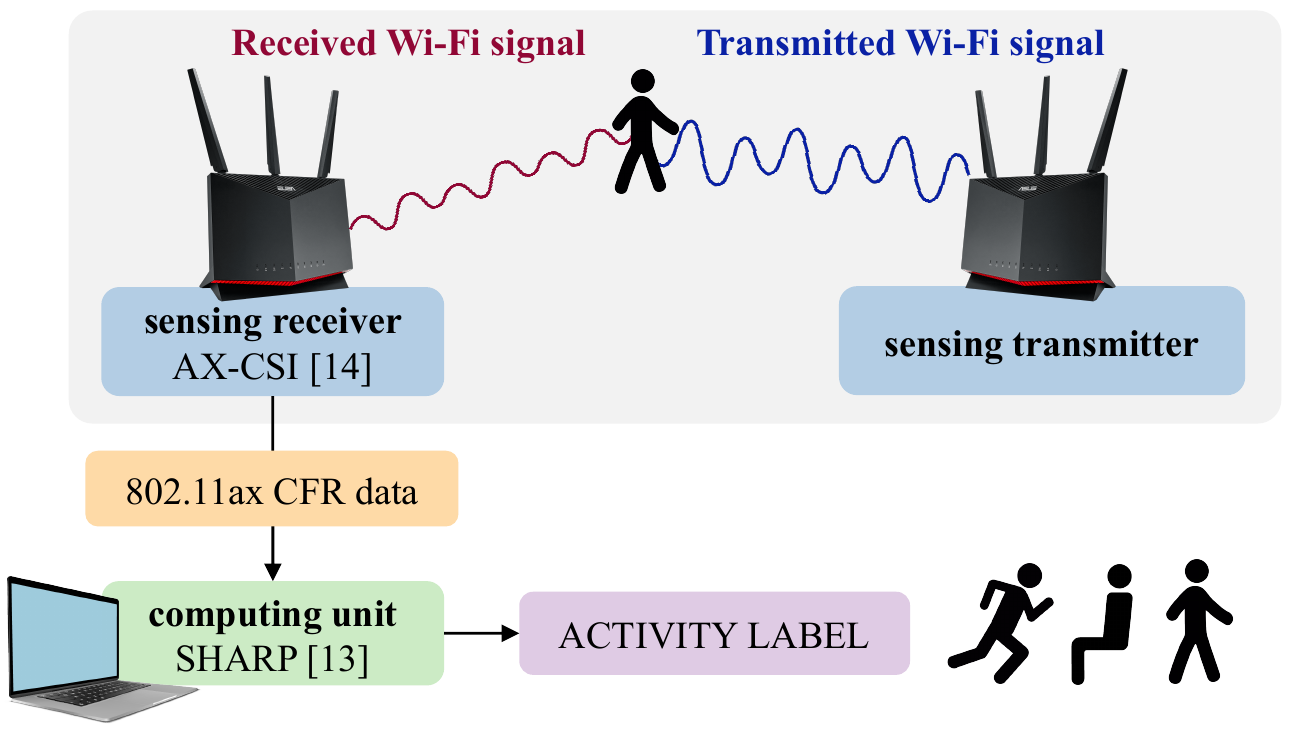}
        \setlength\abovecaptionskip{-.5cm}
    \caption{Experimental setup for sensing data collection and processing.\vspace{-0.4cm}}
    \label{fig:setup}
\end{figure}

To answer this question, we analyze the impact of the sensing bandwidth and the channel sampling period on the classification accuracy. We focus on the human activity recognition task considering SHARP, the state-of-the-art algorithm proposed in~\cite{Meneghello2022SHARP}. We collected a completely novel dataset -- which we pledge to share with the community -- entailing IEEE 802.11 channel data captured in an indoor environment.\footnote{\url{https://github.com/francescamen/SHARPax}} 
Notice that IEEE 802.11bf devices are currently unavailable in the market as they are expected to be commercialized by 2024. Moreover, as discussed above, the main new feature of 802.11bf is to unveil sensing primitives while it is not expected to introduce new transmission schemes that are left to other amendments. In turn, we considered IEEE 802.11ax devices as they implement the latest 802.11 standard release that is currently replacing the majority of Wi-Fi deployments.
To the best of our knowledge, our dataset represents the first collection of 802.11ax \gls{csi} data from commercial devices. In Fig.~\ref{fig:setup} we depicted the network setup together with a summary of the sensing data collection and processing steps followed for the evaluation, as detailed next.

\smallskip
\noindent\textbf{Experimental network setup.} We set up an IEEE 802.11ax network with two Asus RT-AX86U Wi-Fi \glspl{ap}. The network has been deployed in a house corridor by placing the routers along the two long edges, spaced apart by 4~m. The devices exchanged Wi-Fi data over the IEEE 802.11ax channel number 157 using the \gls{ofdma} resource unit RU1-996, i.e., with a bandwidth of 80~MHz and 996 sub-channels.

\smallskip
\noindent\textbf{\Gls{cfr} data collection.} We used the \mbox{AX-CSI} tool to obtain the \gls{cfr} for each packet collected by the receiver~\cite{Gringoli2021ax}. We considered an inter-packet distance of $T_c =$~7.5~ms, being reasonable for sensing applications. We asked a volunteer to perform three activities, i.e., walking and running around the room, and staying in place. We also added an ``empty room'' class, for a total of four classes. For each class, data from four different campaigns -- lasting two minutes each -- were collected. Note that we focus on a limited set of activities and a single subject as we are mainly concerned with studying how a sensing system behaves when changing some communications parameters rather than proposing new sensing strategies. We refer the reader to \cite{Meneghello2022SHARP} for additional evaluations involving more subjects and activities.

\smallskip
\noindent\textbf{\Gls{cfr} data processing.} The \gls{cfr} phase offsets associated with hardware imperfections were corrected using the approach developed in~\cite{Meneghello2022SHARP}. Hence, Doppler vectors were computed every time a new measurement was obtained at the receiver considering a channel observation window of 25 channel readings (the current measurement together with the 24 previous ones), and averaging over the available OFDM sub-channels (see~\cite{Meneghello2022SHARP}). The \gls{dnn} in~\cite{Meneghello2022SHARP} was trained as a four classes classifier. The \gls{dnn} took as input $N =$~256 consecutive Doppler vectors at a time to estimate whether the person was present in the room and, in case, which activity they performed. Once trained, the \gls{dnn} was used to predict the classes on \gls{cfr} data never considered during training, thus allowing for a fair evaluation of the sensing performance. %(around 1.92~s)

\begin{figure}[!t]
    \centering
    \includegraphics[width=1\columnwidth]{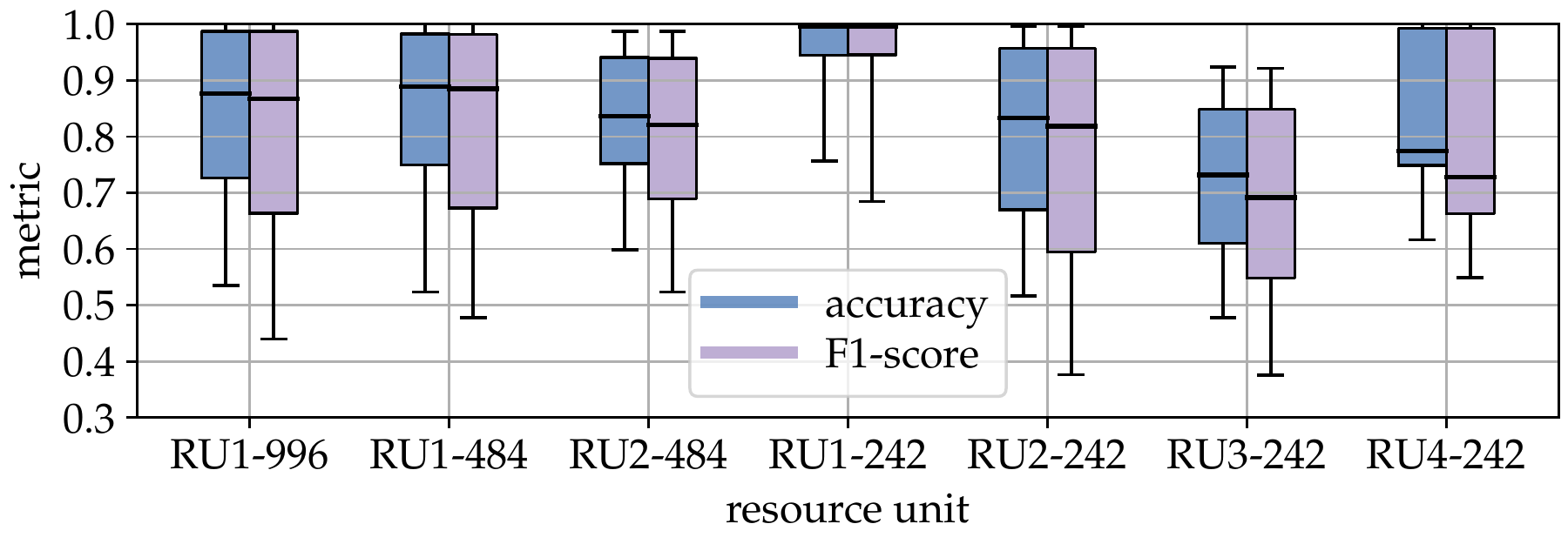}
        \setlength\abovecaptionskip{-.5cm}
    \caption{Average accuracy and F1-score with different \gls{ofdma} \glspl{ru}.\vspace{-0.4cm}}
    \label{fig:change_RU}
\end{figure}

\begin{figure}[!t]
    \centering
    \includegraphics[width=1\columnwidth]{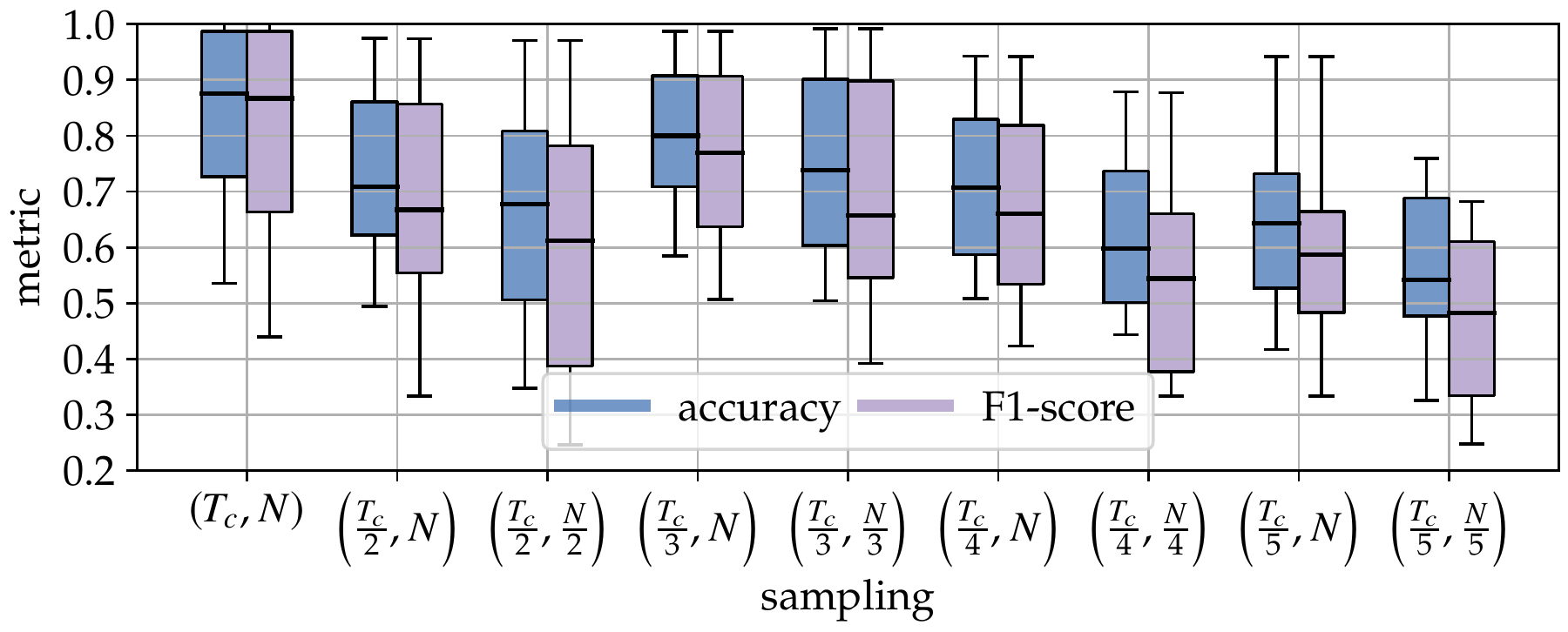}
    	\setlength\abovecaptionskip{-.5cm}
    \caption{Average accuracy and F1-score considering different sampling periods and number of Wi-Fi channel readings used as input for the activity classifier.\vspace{-0.4cm}}
    \label{fig:change_sampl}
\end{figure}

\smallskip
\noindent\textbf{Performance evaluation.} A four-fold cross-validation mechanism has been used, with two campaigns used for training, one for validation, and the remainder for testing. Nine different validation rounds were performed, for a total of 108 evaluation sets. The statistics of the accuracy and F1-score averaged over the 108 tests and the four classes are reported in Figs.~\ref{fig:change_RU}-\ref{fig:change_sampl}. The bars cover the 25-75 percentile interval, the horizontal line within each bar represents the median value, and the whiskers span over the 5-95 percentile interval. 

Fig.~\ref{fig:change_RU} shows the sensing results considering seven different \gls{ofdma} \glspl{ru} as specified by the 802.11ax standard. This allows evaluating how the sensing performance changes when changing the number of \gls{ofdma} sub-channels, and, in turn, the sensing bandwidth. The \glspl{ru} are identified by two numbers where the one after the dash indicates the number of sub-channels, i.e., 996, 484, or 242 for respectively 80~MHz, 40~MHz, and 20~MHz \gls{ru} bandwidth. The number before the dash indicates which of the \glspl{ru} characterized by the same number of sub-channels is considered, i.e., 1, 2, 3, or 4, starting from lower frequency sub-channels to higher frequency ones. The results indicate that there is not a clear link between the number of sub-channels leveraged for sensing and the sensing accuracy. This suggests that -- more than blindly relying on higher bandwidths -- the design of sensing applications should consider properly selecting the sub-channels that are the best for sensing purposes based on some architecture-defined metrics. The higher the number of sub-channels, the more choices are available for the selection process.

In Fig.~\ref{fig:change_sampl} we evaluate the impact of the sampling period on the sensing performance. Each evaluation has been performed by re-sampling the sensing data at RU1-996 considering sampling periods of $T_c/2$, $T_c/3$, $T_c/4$, and $T_c/5$. We also evaluate the impact of changing the number of Doppler vectors used as input for the neural network accordingly to the sub-sampling operations, i.e., $N$, $N/2$, $N/3$, $N/4$, and $N/5$. The first group of bars refers to the reference metrics, i.e., without sub-sampling. We notice that the sensing performance decreases when sub-sampling the signal, even if there is not a clear trend as $T_c/3$ offers better performance than $T_c/2$. Therefore, the sampling period should be properly evaluated for each sensing design. %\vspace{-0.2cm}

\begin{figure*}[t]
    \centering
    \includegraphics[width=1.8\columnwidth]{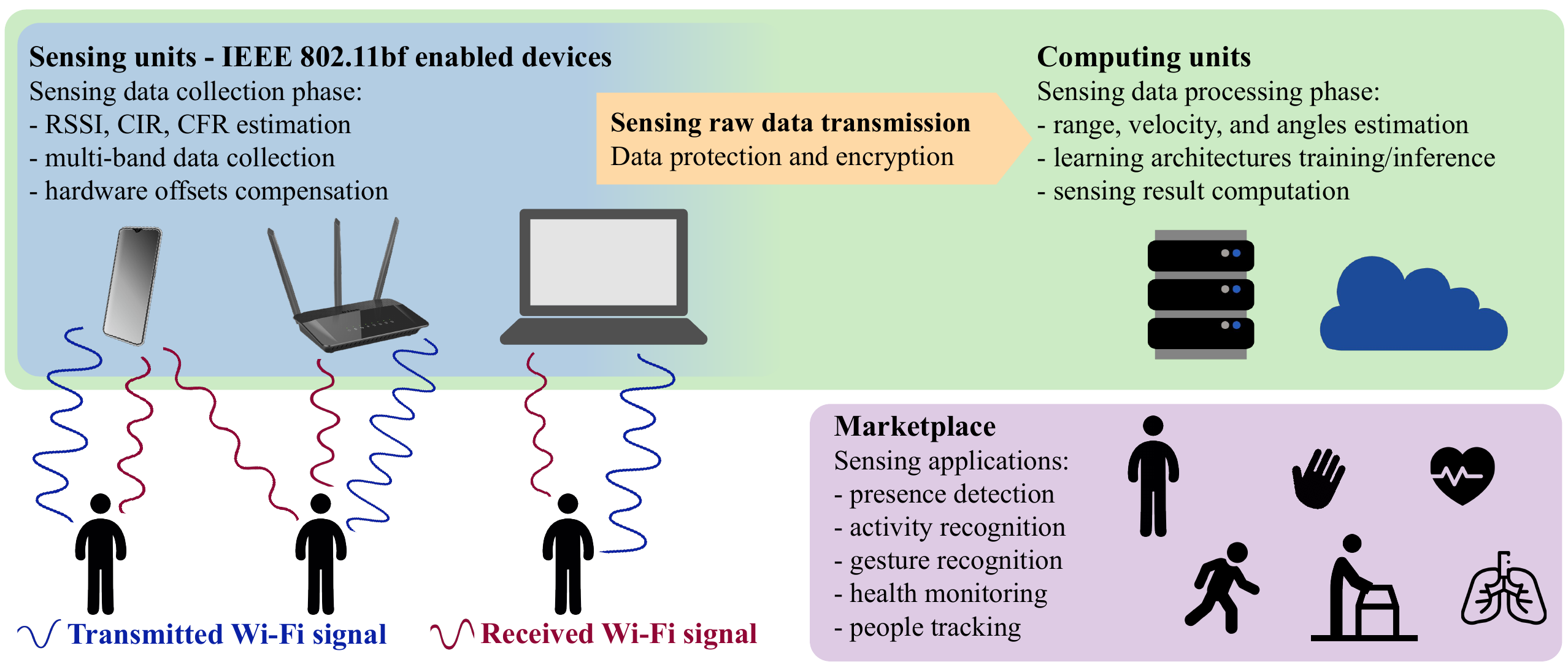}
    \caption{Integration of sensing in Wi-Fi networks. Channel data are collected by the sensing units. Hence, the sensing application is executed on the computing units. Sensing applications can be downloaded from a marketplace.\vspace{-0.5cm}}
    \label{fig:wifi_sensing}
\end{figure*}

\section{Is everything ready? Research Challenges about Integrating Sensing into Wi-Fi Networks}\label{sec:challenges}

Although the  community is actively defining proper \gls{phy}/\gls{mac} layer modifications to enable sensing, it is not clear how communications, computation and sensing services will be highly intertwined. To bridge this gap, we provide an overview of the main research challenges to ISAC in Wi-Fi networks. \vspace{-0.3cm}

\subsection{Data Collection, Transmission and Processing}

\noindent\textbf{Data collection.} Either the Wi-Fi \glspl{ap} or devices such as smartphones, tablets, and laptops, i.e., \glspl{nap}, can gather sensing data (see Fig.~\ref{fig:wifi_sensing} on the left). The device where to execute this phase should be selected based on the required accuracy and Wi-Fi device manufacturers will need to properly consider the sensing needs during the design phases. For example, the antenna placement should be reconsidered as external antennas provide better \gls{snr}, and equally spaced antennas ease the computation of the \gls{aoa} to estimate the position of targets~\cite{Liu2022Integrated}.

\smallskip
\noindent\textbf{Data processing.} For this phase, \mbox{Wi-Fi} \glspl{ap}, \glspl{nap}, and ad-hoc edge devices may serve as computing units. Alternatively, the processing can be offloaded to cloud services (see Fig.~\ref{fig:wifi_sensing} on the right). The choice should be guided by the needed computing power and the time sensitivity of the sensing application. In general, learning-based or hybrid approaches require higher computing power due to the long training process. In this respect, the training is expected to be performed either by the application vendors or demanded to the final users. In the former case, the data is collected, processed, and stored only by the application provider thus the user is not required to collect data for training. This approach is the most convenient from a user privacy perspective. However, it may lead to decreased sensing performance as sensing is actually performed in a different scenario than the ones considered at training. The latter approach consists in providing the user with the sole learning-based architecture that will be trained with user-specific data collected on the final deployment. While this strategy would be the best in terms of the accuracy of the trained algorithm, it may be of difficult applicability as the system would not be plug-and-play. As a tradeoff between the two approaches, few shots adaptation and continual learning algorithms can be considered, and the adaptation can be performed both on the local computing facilities or remotely on the cloud managed by the vendor. The inference phase requires less computing power but still needs memory support to save the learned parameters. To this end, strategies for resource-constrained devices, such as Wi-Fi \glspl{ap} and \glspl{nap}, are being developed. 
Overall, we expect that both on-site and remote computing will be available, and that end users will have access to a marketplace where to download sensing applications for their devices. Each application will have some requirements in terms of sensing data collection and support for computation, and different versions would be made available to provide broad support. Wi-Fi \gls{ap} will probability be provided with some basic sensing features already included, with the possibility to install additional tools depending on the resource availability.

\smallskip
\noindent\textbf{Data transmission.} Depending on where the sensing data collection and the processing phases are executed, the sensing data may need to be transmitted from the sensing data collector to other local or remote entities that manage the processing, as depicted in Fig.~\ref{fig:wifi_sensing}. Such data transmission makes it essential to integrate some data protection and encryption strategies to prevent adversarial attacks against the sensing service. In this respect, IEEE 802.11bf introduces the protected management frames for the sensing measurement report transmission. Moreover, when data is transmitted to the cloud, some techniques should be applied to anonymize the information and prevent possible privacy issues and data leakages. %\vspace{-0.3cm}

\subsection{Sensing Security and Privacy}\label{subsec:security}
\begin{figure}[!h]
    \centering
    \includegraphics[width=0.9\columnwidth]{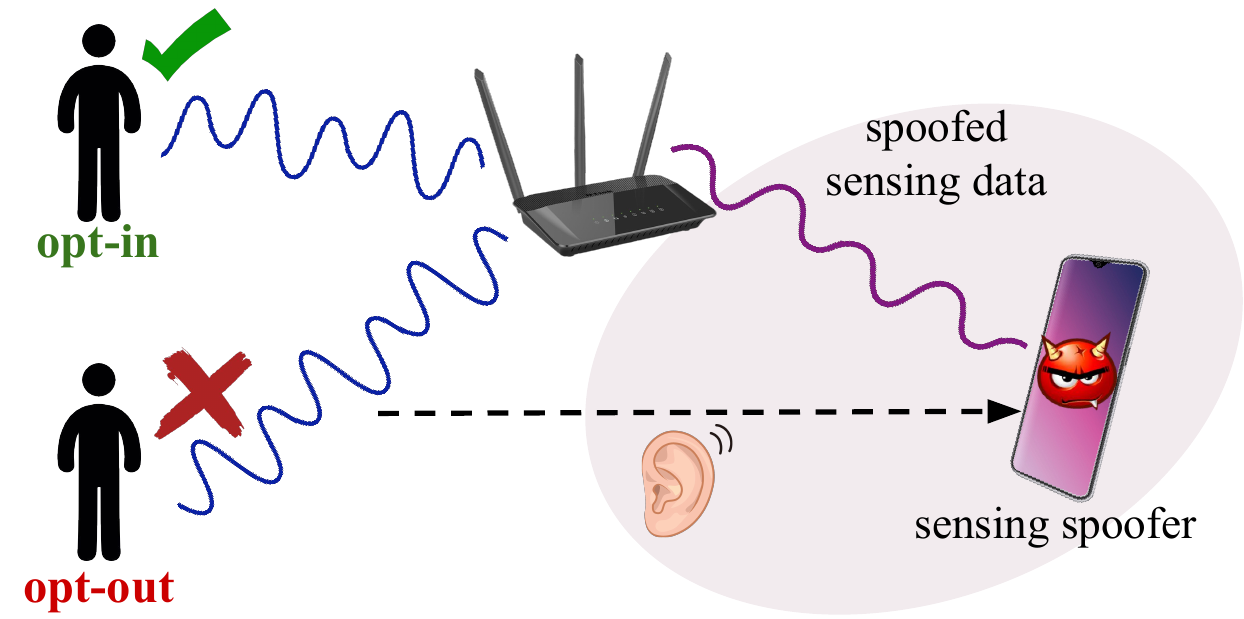}
        \setlength\abovecaptionskip{-.1cm}
    \caption{Sensing security and privacy.}
    \label{fig:sens_challenges_a} 
    \vspace{-0.3cm}
\end{figure}

The pervasiveness of sensing into our everyday lives will necessarily elicit security and privacy concerns. Given the broadcast nature of the wireless channel, a malicious eavesdropper could easily capture the \gls{csi} reports and track the user's activity without authorization. Worse yet, end-users may not even realize they are under attack when using radio-frequency-based monitoring solutions. In fact, with respect to cameras, wireless sensing applications also work in the dark, with smoke or dust in the environment, and when obstacles -- e.g., walls, furniture -- are between the sensing device and the subject (operating on the sub-7 GHz bands). 
However, as yet, research and development efforts have been focused on improving the classification accuracy of the phenomena being monitored, with little regard to security and privacy issues. 
To address this point, the first important aspect is the development of \gls{dnn}-based Wi-Fi sensing systems robust to adversarial machine learning techniques. Moreover, individuals should be provided the opportunity to \textit{opt out} of sensing services, as depicted on the left side of Fig.~\ref{fig:sens_challenges_a}. This would require the widespread introduction of reliable sensing algorithms for subject identification. Although some techniques have been proposed \cite{ma2019wifi}, it is unclear whether they are resilient to malicious users actively trying to impersonate other users, as shown on the right side of Fig.~\ref{fig:sens_challenges_a}, or adverse channel conditions, i.e., presence of noise and interference from other technologies. Identification techniques should also be tested against adversaries, either through active techniques, i.e., a device carefully jamming the sensing activity, or passive techniques, i.e., materials shielding and/or deflecting the \mbox{Wi-Fi} radiation. 
Another issue arises when the malicious entity estimates the \gls{csi} and performs sensing on ongoing Wi-Fi traffic. Here, a possible solution is to encrypt the training fields of the data packets so that only trusted devices can retrieve them and estimate the \gls{csi}. This option was already adopted in IEEE 802.11az to protect the location/ranging information from potential eavesdroppers. \vspace{-0.3cm}

\subsection{Cooperative and Multi-band Sensing}
\begin{figure}[!h]
    \centering
    \includegraphics[width=0.85\columnwidth]{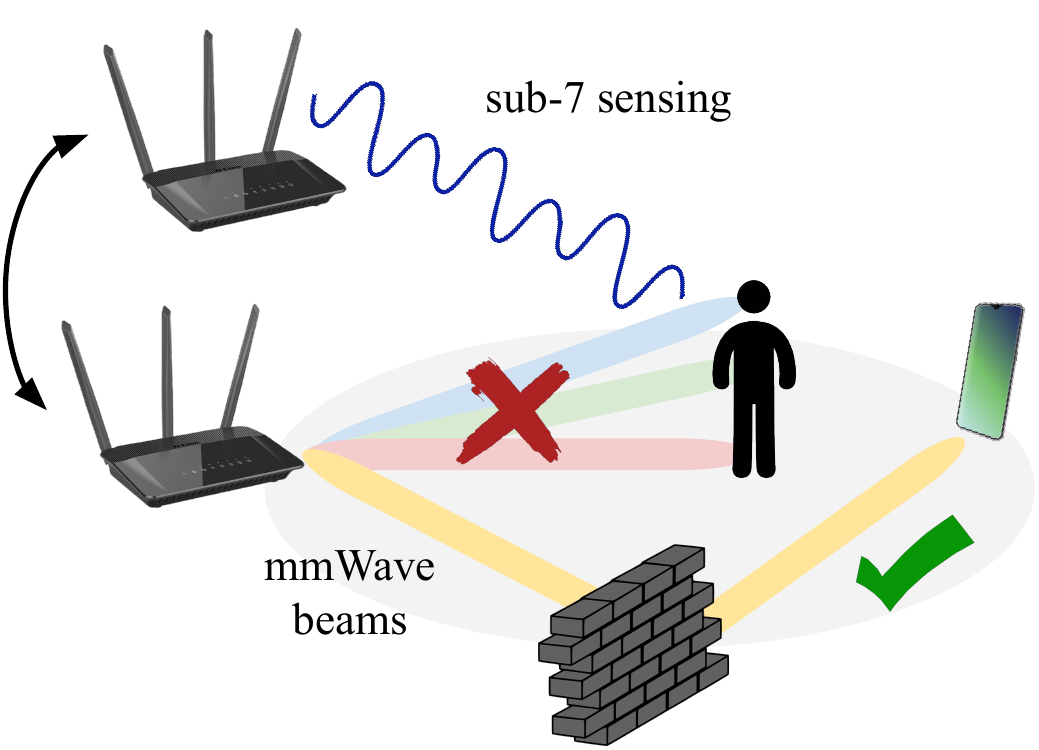}\vspace{-0.2cm}
    \caption{Multi-band cooperative sensing-aided Wi-Fi Systems.\vspace{-0.5cm}}
    \label{fig:sens_challenges_b} 
\end{figure}

Cooperative and multi-band sensing will provide a unique opportunity to not only boost the sensing accuracy, but also to leverage the increased location awareness of blockages to design intelligent sensing-aided Wi-Fi communications that will ameliorate the performance of \gls{mmw} Wi-Fi links. For example, understanding the size and movement of blocking entities through sub-7 \gls{csi} reports could guide beam selection in the \gls{mmw} link, as shown in Fig.~\ref{fig:sens_challenges_b}. By the same token, understanding the location of a \gls{nap} by using sub-7 sensing can help reduce the overhead associated with beam scanning and alignment. 
A key challenge will be to coordinate time-sensitive cooperative sensing operations among multiple Wi-Fi devices in different spectrum bands. Indeed, communication-related sensing will be extremely time-sensitive for different safety-critical applications such as autonomous driving and telemedicine, or for virtual/augmented/mixed reality and holography for entertainment and remote working. These applications require the sensing information to be available at the communication end-point within milliseconds from the acquisition. To this end, a possible strategy could be to introduce control channels in the sub-7 band exclusively dedicated to the coordination of low-latency cooperative sensing operations.\vspace{-0.5cm} 

\subsection{Sensing in Spectrum-Sharing Bands}

From IEEE 802.11ax onward, \mbox{Wi-Fi} devices will share the spectrum with incumbents in the 6 GHz band, such as licensed point-to-point and satellite services, as well as other license-exempt ultra-wideband systems and 5G NR-Unlicensed. To protect incumbent services, license-exempt devices operate under restrictions such as maximum emitted power and indoor-only operation. Given the intense spectrum sharing in the 6~GHz band, further investigations should address how to make sensing robust to interference.\vspace{-0.3cm}

\subsection{Integrating Sensing and Communications}

To make communication and sensing services coexist in Wi-Fi networks, sensing transmissions -- i.e., performed to obtain channel estimates -- could be ``piggybacked'' into data packets to avoid decreasing the communication throughput. However, data packets may be subject to significant interference in the 6~GHz band, which may be tolerable for data recovery but intolerable from a sensing perspective. Therefore, a core issue is to determine the optimal trade-off between making reserved use of the spectrum for sensing operations and piggybacking sensing into data packets. Similar to multi-band sensing, dedicated channels could be used to improve sensing performance without a significant decrease in system throughput.\vspace{-0.2cm}

\section{Concluding Remarks}

Sensing services are expected to be implemented within \mbox{Wi-Fi} networks by 2024 through the release of the IEEE 802.11bf standard. Researchers are currently working on two parallel directions that will enable  integrating sensing into \mbox{Wi-Fi} networks. The Wi-Fi technological peculiarities leveraged for sensing purposes are detailed in this article, together with the approaches to developing Wi-Fi sensing algorithms. We included practical lessons learned from experimental evaluations with commercial devices and an overview of the open research challenges. Overall, we trust that our contribution will provide a comprehensive overview of the opportunities and challenges of Wi-Fi sensing.\vspace{-0.2cm}

\section*{Acknowledgment}

This material is based upon work supported by the National Science Foundation under Grant No. CNS-2134973 and CNS-2120447. 
The work was also partially supported by the European Union under the Italian National Recovery and Resilience Plan (NRRP) of NextGenerationEU, partnership on ``Telecommunications of the Future'' (PE00000001 - program ``RESTART''), and by the Fulbright Schuman Program, administered by the Fulbright Commission in Brussels and jointly financed by the U.S. Department of State, and the Directorate-General for Education, Youth, Sport and Culture (DG.EAC) of the European Commission. The views and opinions expressed in this work are those of the authors and do not necessarily reflect those of the funding institutions.\vspace{-0.2cm}

\footnotesize
\bibliographystyle{IEEEtran}
\bibliography{biblio}

\begin{IEEEbiographynophoto}
{Francesca Meneghello} (M'19) is an Assistant Professor at the University of Padova (Italy), Department of Information Engineering. Her research interests include radio frequency sensing and wireless networks.
\end{IEEEbiographynophoto}
\vspace{-1cm}
\begin{IEEEbiographynophoto}
{Cheng Chen} is a Software Research Engineer/Scientist at Intel Corporation. He received the Ph.D. degree from Northwestern University, Evanston, IL in 2016.
\end{IEEEbiographynophoto}
\vspace{-1cm}
\begin{IEEEbiographynophoto}
{Carlos Cordeiro} is the wireless CTO at Intel's client group and an Intel Fellow. He serves as the chair of the Wi-Fi Alliance Board of Directors, the associate editor-in-chief of the IEEE Communications Standards Magazine. He is an IEEE Fellow.
\end{IEEEbiographynophoto}
\vspace{-1cm}
\begin{IEEEbiographynophoto}
{Francesco Restuccia} (SM'21) is an Assistant Professor of Electrical and Computer Engineering at Northeastern University, United States. His research interests lie at the intersection of wireless networks, artificial intelligence and embedded systems. He is a Senior Member of IEEE.
\end{IEEEbiographynophoto}

\vspace{-1cm}

\vfill

\end{document}